\documentclass{osa-article}

\journal{oe}

\usepackage{color}
\articletype{Research Article}

\begin{document}

\title{Ultraprecise Rydberg atomic localization using optical vortices}

\author{Ning Jia,\authormark{1} Teodora Kirova,\authormark{4} Gediminas Juzeli$\overline{u}$nas,\authormark{3} Hamid Reza Hamedi,\authormark{3,*} and Jing Qian\authormark{2,$\dagger$}}

\address{\authormark{1} The Public Experimental Center, University of Shanghai for Science and Technology, Shanghai 200093, China\\
\authormark{2}State Key Laboratory of Precision Spectroscopy, Quantum Institute for Light and Atoms,
Department of Physics, School of Physics and Electronic Science,
East China Normal University, Shanghai 200062, China\\
\authormark{3}Institute of Theoretical Physics and Astronomy, Vilnius University, LT-10257, Lithuania\\
\authormark{4}Institute of Atomic Physics and Spectroscopy, University of Latvia, LV-1004, Latvia}

\email{\authormark{*} hamid.hamedi@tfai.vu.lt} 
\email{\authormark{$\dagger$}jqian1982@gmail.com}



\begin{abstract}
We propose a robust localization of the highly-excited Rydberg atoms interacting with doughnut-shaped optical vortices. Compared with the earlier standing-wave (SW)-based localization methods, a vortex beam can provide an ultraprecise two-dimensional localization solely in the zero-intensity center, within a confined excitation region down to the nanometer scale. We show that the presence of the Rydberg-Rydberg interaction permits counter-intuitively much stronger confinement towards a high spatial resolution when it is partially compensated by a suitable detuning. In addition, applying an auxiliary SW modulation to the two-photon detuning allows a three-dimensional confinement of Rydberg atoms. In this case, the vortex field provides a transverse confinement, while the SW modulation of the two-photon detuning localizes the Rydberg atoms longitudinally. To develop a new subwavelength localization technique, our results pave one-step closer to reduce excitation volumes to the level of a few nanometers, representing a feasible implementation for the future experimental applications.
\end{abstract}

\section{Introduction}

Recent years have seen a vast progress in the precise localization of atoms, with potential applications in the fundamental and applied science. Some important examples are precise addressing of ultracold atoms in optical lattices \cite{PRL150501,Nelson2007}, patterning of Bose-Einstein condensates (BECs) \cite{pra061601,pra053638}, optical lithography \cite{lithography2005} or fluorescence microscopy \cite{fluorescence2007}. The diffraction limit, however, is a barrier to the possible resolution. For example, in an optical microscope, the highest achievable point-to-point resolution is limited by diffraction. The diffraction restricts the  ability of optical instruments to distinguish between two objects separated by a distance smaller than a half of the wavelength of light employed to image the sample.

Coherent-adiabatic light-matter interactions provide some ways to overcome the diffraction limit by reducing the excitation volume in atom-light coupling schemes. The concepts are based on spatially modulated dark-states created by the Electromagnetically Induced Transparency (EIT) \cite{EIT2006,EIT2007,EIT2008,EIT2014} or the Coherent Population Trapping (CPT) \cite{CPT1}. The space-dependent interaction between the light field and the atomic internal states is produced by a standing-wave (SW) field. New schemes have been proposed for the SW localization by applying different measuring ways, {\it e.g.} absorption spectrum \cite{absorb3,absorb4}, level population  \cite{population1,population2,population3,population4}, spontaneous emission  \cite{emmision1,emmision2,emmision3}, or adopting complex energy-level structures  \cite{absorb1,absorb2}. Beyond the theory proposals, there are only a few experiments on EIT-based localization \cite{expr2011,expr2013,EIT2015},
demonstrating an atomic localization to regions of 60 nm, {\it i.e.} 13 times smaller than the wavelength of incident light \cite{EIT2015}.
The precision level better than 30nm has been achieved in imaging of molecules and biological dynamics \cite{bio,bio1}. Yet improving the resolution of subwavelength atomic localization down to the range of a single nanometer remains an important challenge.

A localization protocol with SW produces a periodic pattern of tightly localized regions. This was fine for the first experimental demonstrations \cite{expr2011,expr2013,EIT2015}, but it is not appropriate for applications, which usually require single excitation regions. 
 On the other hand, in the search for systems suitable for quantum information and precision measurement \cite{information2010,gate2000,gate2017,simu2019}, Rydberg atom has emerged as one of the favorites mainly due to its strong long-range interaction that blocks the possibility of multiple excitations \cite{Rb1,Rb2,Rb3,Rb4,Rb5}. However, this superiority also gives rise to a poor quality of localizing Rydberg atoms because it is difficult to confine them in a small region with a high density. Hence, it is still under way to precisely localize highly-excited Rydberg atoms using currently available experimental techniques. Very recently we have studied the localization of  Rydberg atoms via SW beams that produce a periodic pattern of tightly localized atoms in one dimension\cite{Kirova:20}, however, the current work provides a progress towards single site confinement of atoms.

In order to find atoms in a single excitation region, in the present work, we propose and analyze a theoretical scheme of combining Rydberg atoms with a special space-dependent doughnut-shaped beams which carry an orbital angular momentum (OAM)  \cite{a1,Babiker2018}. In contrast to the earlier SW-based localization protocols, a doughnut beam geometry makes it possible to detect atoms in a single spatial region with a 100$\%$ probability, where any fluctuations from the laser noise can be largely suppressed. It should be noted that the localized excitation of a four-level atom to a highly excited Rydberg state has been theoretically investigated in \cite{Mashhadi} by considering the Laguerre-Gaussian (LG) beam spatial features. Yet, the dipole-dipole interaction between the Rydberg atoms (which may induce the blockade) has been neglected, allowing in \cite{Mashhadi} to focus on the single atom excitation mechanism.
However, the work presented in this paper is completely different.
It is shown here that the strong Rydberg-Rydberg interaction can be partially compensated by a suitable detuning, and a two-dimension (2D) transverse confinement can be achieved with a localization precision $\sim$11nm. To image the Rydberg atoms in the three-dimensional (3D) space, in addition to the vortex beam, we employ an auxiliary SW modulation to the two-photon detuning. Such a SW modulation yields the longitudinal placement of Rydberg atoms along the propagation direction of the probe beam, while the atomic excitation is spatially confined in a transverse plane due to application of vortex beams. In addition, we explore an experimental implementation of our setup under realistic parameters, strongly supporting the preservation of the robustness of our 3D Rydberg localization protocol, even under the influences of non-negligible random intensity noise and frequency noise.

\section{Theoretical Formulation}
\begin{figure}
\centering
\includegraphics[width=0.49\textwidth]{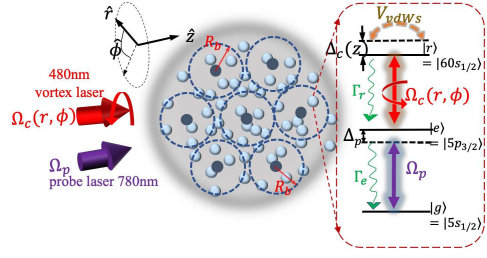} %
\caption{Schematic diagram for a collection of Rydberg superatoms interacting with a TW field $\Omega_p$ as well as a LG field $\Omega_c(r,\phi)$. Both beams are propagating along the same direction $\hat{z}$. The concept of superatom is that, within the blockade radius $R_b$, only one of the atoms can obtain one excitation to the uppermost Rydberg state.
Inset: Level structure of each atom with states $|g\rangle$, $|e\rangle$, $|r\rangle$ denoting the ground, intermediate and Rydberg states, which enables the transitions of $\left|g\right\rangle \leftrightarrow\left|e\right\rangle$ and $\left|e\right\rangle \leftrightarrow\left|r\right\rangle$. $V_{vdWs}$ stands for the intrinsic $ns-ns$ type $vdWs$ Rydberg interaction between the unique excited atoms of adjacent superatoms. Other parameters are described in the text.}
\label{model}
\end{figure}

Let us consider an ensemble of atoms characterized by a typical three-level ladder configuration of energy levels as shown in  Fig. \ref{model}. For each atom, states $\{|g\rangle,|e\rangle,|r\rangle\}$ represent the ground, the excited and the highly excited Rydberg states, respectively. The transition between $|g\rangle$ and $|e\rangle$ is induced by a traveling-wave (TW) field characterised by the Rabi-frequency $\Omega_p$ and the frequency detuned by $\Delta_p$ from $|e\rangle$. The upper transition between $|e\rangle$ and $|r\rangle$ is driven by a vortex control field $\Omega_c(r,\phi)$ , which is detuned by $\Delta_c$ with respect to $|r\rangle$. The control field detuning $\Delta_c(z)$ can be adjusted to be $z$-dependent enabling an auxiliary spatial confinement along the $z$-axis.

We assume a frozen-atom limit where the atomic center-of-mass motion is negligible due to the fast operation of experiments $\sim\mu$s  \cite{Frozen1,Frozen2,Frozen3}. Therefore, applying the rotating-wave approximation, the Hamiltonian is given by ($\hbar=1$)
\begin{equation}
\mathcal{H}=\mathcal{H}_a+\mathcal{V}_{af}+\mathcal{V}_{vdWs},
\label{haml}
\end{equation}
where
\begin{eqnarray}
\mathcal{H}_a&=&- \sum_{j}^{N}\left[\Delta_p\sigma_{ee}^j+(\Delta_c+\Delta_p)\sigma_{rr}^j \right],  \\
\mathcal{V}_{af}&=&-\sum_{j}^{N}\left[\Omega_p \sigma_{eg}^{j}+\Omega_c \sigma_{re}^{j}+H.c.\right], \\
\mathcal{V}_{vdWs}&=& \sum_{j<m}^{N} \frac{C_6}{\left|\mathbf{r}_j-\mathbf{r}_m\right|^6}\sigma_{rr}^{j}\otimes\sigma_{rr}^{m}
\label{H}
\end{eqnarray}
are the unperturbed atomic Hamiltonian $\mathcal{H}_a$, the atom-field interaction $\mathcal{V}_{af}$ and the internuclear van der Waals({\it vdWs})-type interaction $\mathcal{V}_{vdWs}$, respectively. Note that the electric dipole approximation (EDA) is adopted while deriving Eqs.(\ref{haml}-\ref{H}). In general the EDA is applicable if $kr\ll1$, with $k=2\pi/\lambda$ representing the wavevector of field and $r$ the atomic size. For a higher Rydberg level this condition is hard to meet automatically.
 Due to the special property of the LG beam the Rydberg electron may "see" the light intensity variation, {\it i.e.}
different parts of the atom interact with different electric fields. The resulting quadrupole Rabi frequency related to the LG beam and the Rydberg state can cause additional OAM exchange covering the center-of-mass and internal motions \cite{Rodrigues, Mukherjee}. By approximately neglecting the OAM transfer between light and atomic internal degrees we focus on the intensity variation of the LG beam and will leave the interesting physics of higher-order effects for a future work.

As for the $j$th atom, $\sigma_{\alpha\beta}^{j}=\left|\alpha\right\rangle \left\langle \beta\right|_{j}$ is the transition ($\alpha\neq\beta$) or projection ($\alpha=\beta$) operator, while $C_6$ denotes the {\it vdWs} coefficient which depends on $|r\rangle$. Under the mean-field treatment \cite{mean_field}, one can safely replace $\mathcal{V}_{vdWs}$ with  $\sum_{j}^{N}\sigma_{rr}^{j}\sum_{m\neq j}\frac{C_6}{\left|\mathbf{r}_j-\mathbf{r}_m\right|^6}\sigma_{rr}^{m}$. The mean-field approximation neglects the correlations of Rydberg-state atoms within a single superatom, which is enabled by strong Rydberg blockade that prevents the excitation of a second atom within a superatom. The correlation can be incorporated into the expression of $s$ [Eq.\ref{s}] via a short-range cutoff to the spatial integral for the Rydberg interaction. For short excitation times, this method yields good agreement with experiments\cite{022709,PRL250601}. The time evolution of operator $\sigma_{\alpha\beta}^j(t)$ of the $j$th atom is governed by
\begin{eqnarray}
\label{Bloch_eqs}
\dot{\sigma}_{gg}^j &=& \Gamma_e \sigma_{ee}^j  -2 \text{Im}\left(\Omega_p^{*}\sigma_{ge}^j  \right), \nonumber\\
\dot{\sigma}_{ee}^j &=&\Gamma_r\sigma_{rr}^j-\Gamma_{e}\sigma_{ee}^j-2\text{Im}\left(\Omega_c^{*}\sigma_{er}^j\right)+2\text{Im}\left(\Omega_p^{*}\sigma_{ge}^j\right), \nonumber \\
\dot{\sigma}_{ge}^j &=&\left(i\Delta_p-\gamma_{ge}\right)  \sigma_{ge}^j+i\left[\Omega_c^{*}\sigma_{gr}^j -\Omega_{p}(\sigma_{ee}^j-\sigma_{gg}^j)\right], \label{motion} \\
\dot{\sigma}_{er}^j &=&\left[i \left(\Delta_c-s\right)-\gamma_{er}\right]  \sigma_{er}^j -i \left[\Omega_{p}^{*}\sigma_{gr}^j  +\Omega_c(\sigma_{rr}^j-\sigma_{ee}^j)\right],  \nonumber\\
\dot{\sigma}_{gr}^j &=&\left[i\left(\Delta_{p}+\Delta_{c}-s\right)-\gamma_{gr}\right]\sigma_{gr}^j+i\left(\Omega_{c} \sigma_{ge}^j-\Omega_{p} \sigma_{er}^j\right), \nonumber
 \end{eqnarray}
where
\begin{equation}
 s=\sum_{m\neq j}\frac{C_6}{\left|\mathbf{r}_j-\mathbf{r}_m\right|^6 } \sigma_{rr}^{m}
  \label{sseq}
\end{equation}
is the accumulated {\it vdWs}-induced energy shift for the atom $j$ induced by the adjacent Rydberg-state atoms $m$,
and $\gamma_{\alpha\beta}=\left(\Gamma_\alpha+\Gamma_\beta\right)/2$ is the dephasing rate,  with $\alpha,\beta\in(g,e,r)$. 
If the spontaneous decay rates obey the condition $\Gamma_e\gg\Gamma_{r}$, $\Gamma_g\approx0$, one has approximately $\gamma = \gamma_{er}=\gamma_{ge}$, $\Gamma_e=2\gamma$ and $\gamma_{gr}=\Gamma_r  \approx0$. The two-photon detuning is described by $\Delta_p+\Delta_c$. In addition, we will ignore the superscript $j$ for the sake of simplicity. Solving the system of equations (\ref{motion}) under the steady limit ($\dot\sigma_{\alpha\beta}(t)\equiv0$),  one arrives at a steady-state solution $\sigma_{rr}$ indicating the stable population for the state $|r\rangle$:

\begin{equation}
\sigma_{rr}(\boldsymbol{r})=\frac{I_{p}\left(I_{p}+I_{c}\right)}
{\left(I_{p}+I_{c}\right)^{2}-2\Delta_p \left(\Delta_p+\Delta_c-s\right)I_c+\left(\gamma^{2}+\Delta_p^2+2I_{p}\right)\left(\Delta_{p}+\Delta_c-s\right)^{2}}\label{sigmarr}
\end{equation}
with the laser intensity $I_{p\left(c\right)}=\left|\Omega_{p\left(c\right)}\right|^2$. For $s=0$, the solution $\sigma_{rr}$ has a Lorentzian dependence on the two-photon detuning $(\Delta_p+\Delta_c)$, with its half-peak width given by $w = (I_p+I_c)/\sqrt{\gamma^2+\Delta_p^2+2I_p}$ \cite{Dan}.

For estimating the Rydberg-Rydberg interaction between the neighboring excited atoms belonging to different superatoms, the blockade radius is defined by $R_b=(C_6/w)^{1/6}$ if assuming $\hbar w = C_6/R_b^6$ ($\hbar=1$) \cite{Rb5}. Then only one atom can be excited within a single superatom volume $V_b$. It is apparent that the blockade radius $R_b$ is also position-dependent due to the spatial dependence of the control field and hence $w$. The Rydberg interaction $s$ felt by the $j$th excited atom within a single $V_b$ is calculated by integrating over all excitation probabilities from the volume $V\neq V_b$ \cite{022709}
\begin{equation}
 s=\int_{V\neq V_{b}}\frac{C_{6}}{\mathbf{r}^{6}} f_{R}\rho d^{3}\mathbf{r}.  
 \label{s}
\end{equation}
Here $\rho$ is the ground atom density and $\mathbf{r} = |\mathbf{r}_j - \mathbf{r}_m| $ denotes the relative distance. The average Rydberg excitation fraction $f_{R}$ is described by \cite{collective2,s_f0}
\begin{equation}
 f_{R}=\frac{f_0}{1+(N_{sa} -1)f_0} ,
\end{equation}
where the Rydberg population fraction is $f_0 =\sigma_{rr}$ at $s=0$, and $N_{sa}=V_b\rho$ represents the number of atoms in a single superatom. Note that if $N_{sa}=1$ we get $f_{R}=f_0$ meaning that only one atom inside can obtain a determined excitation; otherwise, assuming $N_{sa}\gg1/f_0$, one gets $f_{R}=1/N_{sa}$ indicating that the blocked volume definitely contains one Rydberg excitation and the Rydberg excitation fraction for each atom is $1/N_{sa}$.
The resulting $s$ becomes $s = \int_{V\neq V_{b}}\frac{C_6}{\mathbf{r}^6}\frac{\rho}{N_{sa}}d^3\mathbf{r}$. In appendix A, we discuss the calculation of the shifted energy $s(\mathbf{r}_j)$ with respect to the $j$th atom in details.

\section{Transverse super localization}

\subsection{Atomic spatial resolution}
Let us first consider a perfect antiblockade condition $\Delta_p+\Delta_c-s = 0$, where the two photon detuning $\Delta_p+\Delta_c$ compensates the Rydberg shift $s$. Without loss of generality in what follows we assume $\Delta_p=0$. As a result, $\Delta_c$ actually stands for the two photon detuning and $\Delta_c-s \equiv0$ represents the perfect antiblockade condition. In this case, the population of the Rydberg state given by Eq. (\ref{sigmarr}) takes the form
\begin{equation}
    \sigma_{rr}=\frac{1}{1+\eta}.
    \label{sigma0}
\end{equation}
Here $\eta=I_c/I_p$ represents a relative ratio between the two laser intensities. Equation (\ref{sigma0}) acquires its maximum value in the core of the vortex beam where $\Omega_c=0$ and hence $\eta=0$. Thus, ensuring the perfect antiblockade condition and so long as the steady state is reached, monitoring the population of Rydberg state is a sufficient tool to measure the position of atoms.

Inspired by this, we take the control laser $\Omega_c(r,\phi)$ to be a doughnut-shaped Laguerre-Gaussian (LG) beam of the form  \cite{Hamid100}
\begin{equation}
\Omega_c(r,\phi)=\Omega_{c0}\left(\frac{r}{W_0}\right)^{\left|l\right|}e^{-r^2/W_0^2}e^{i l \phi},
\label{VO}
\end{equation}
with a zero-intensity solely at the beam core $r=0$. Other higher-order vortex fields with more radial nodes may cause multi-site atom localization, and can be considered for a future study Here $\Omega_{c0}$, $W_0$ and $l$ are, respectively, the peak amplitude, the beam waist, and the winding number of the vortex beam, while $(r,\phi)$ are, respectively, the cylindrical radius and the azimuthal angle. Under the perfect antiblockade condition, the system evolves into a position dependent dark state $|D\rangle=\frac{(\Omega_p|r\rangle-\Omega_c(r,\phi)|g\rangle)}{\sqrt{\Omega_p^2+\Omega_c^2(r,\phi)}}$. As the position $r$ changes, the system adiabatically follows such a
dark state. One can see that the dark state reduces to the excited Rydberg state $|r\rangle$ at the core of the vortex beam where $\Omega_c\approx0$. Thus, a smooth adiabatic change of $|D\rangle$ can excite atoms to the Rydberg state as $\Omega_c(r,\phi)$ passes through its zero intensity core, and hence the population of $|r\rangle$ can be tightly localized.
 Besides the probe field denoted by $\Omega_p$ describes a TW propagating along the same direction $\hat{z}$ as the LG laser beam with a constant amplitude $\Omega_{p0}$. Note that the spatial modulation of $\Delta_c(z)$ has been ignored here for accomplishing an ideal 2D localization.

Using \eqref{VO} for $\Omega_{c}(r,\phi)$, the intensity ratio $\eta$ takes the form
\begin{equation}
\eta=\kappa^2 \left(\frac{r}{W_o}\right)^{2\left|l\right|}e^{-2(r/W_0)^2}
\end{equation}
with $\kappa=\Omega_{c0}/\Omega_{p0}$. One gets $\eta=0$ at the beam core corresponding to $r=0$. This yields a perfect confinement with a 100$\%$-probability of finding atoms at the vortex core. 

\begin{figure}
\centering
\includegraphics[width=0.6\textwidth]{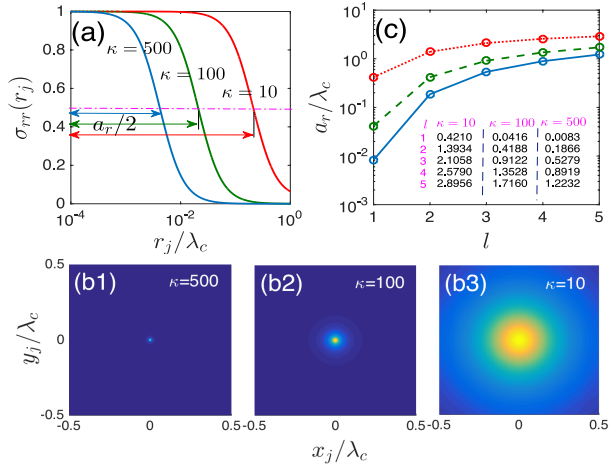} %
\caption{(a) Representation of the steady Rydberg probability $\sigma_{rr}(r_j)$ versus $r_j/\lambda_c$ for $l=1$. The subscript $j$ means for the $j$th Rydberg atom. The full-width at half maximum of excitation where $\sigma_{rr}=0.5$ is defined by $a_r$. Here a half-width $a_r/2$ is labeled. (b1-b3) The 2D plot of atom transverse localization with different $\kappa$ values. (c) The full-width $a_r$ (in unit of $\lambda_c$) with the increase of the winding number $l$. Detailed values $a_r$ (in unit of $\lambda_c$) are shown in the inset. Here the beam waist and the wavelength are $W_0=1\mu$m, $\lambda_c = 480$nm. Cases of $\kappa=10,100,500$ are given by red-dotted, green-dashed and blue-solid curves, respectively.}
\label{width}
\end{figure}

The localization quality depends on a high spatial resolution characterized by a narrow linewidth of steady Rydberg population $\sigma_{rr}(r)$.
A very narrow linewidth indicates that the position of the atoms can be well resolved within a very small excitation volume. For the lowest-order mode of the LG beam with $l=1$, $\eta$ can be expanded around $r=0$ in a Taylor series up to the fourth order, giving
  \begin{equation}
  \eta\approx \kappa^2[(\frac{r}{W_0})^2-2(\frac{r}{W_0})^4].
  \end{equation}
Then the full width at half maximum (FWHM) labeled by $a_r$ of $\sigma_{rr}$ can be approximated as
\begin{equation}
a_r\approx W_0\sqrt{1-\frac{\sqrt{\kappa^2-8}}{\kappa}}.
\label{a_r}
\end{equation}
According to the symmetry, we have $a_r=a_x=a_y$ where $a_x$ and $a_y$ are the FWHM of the excitation along $\hat{x}$ and $\hat{y}$ directions, respectively. From Eq. (\ref{a_r}), it is intuitive that $a_{r}\to0$ only when $\kappa\gg2\sqrt{2}$, indicating a high resolution peak. A very large $\kappa$ is obtained for sufficiently weak probe pulses ($\Omega_{p0}\ll \Omega_{c0}$). As confirmed by Fig.\ref{width}(a), starting from its peak value 1.0, $\sigma_{rr}$ is found to decrease rapidly as $r$ grows, allowing an ultra-precise
confinement of center-of-mass of Rydberg atoms as long as a sufficiently large $\kappa$ is adopted.  However, for a larger value of $\kappa$, the time required for the steady state to be formed is longer, as $\Omega_{p0}$ is weaker. In contrast, a smaller value of $\kappa$ {\it e.g.} $\kappa=10$, would cause a poor resolution although it may speed up the time to reach the steady state. The full numerical results by solving equations (\ref{motion}) confirm the analytical predictions. The relation between steady time and resolution for different $\kappa$ values will be discussed in section 5.2. Figure \ref{width}(b1-b3) shows the 2D imaging of the atomic localization with different $\kappa$ values. It is clear that the localization quality becomes worse as $\kappa$ decreases, indicating the importance of a sufficiently weak probe field for a tight 2D confinement of atoms.

The dependence of $a_r$ on the OAM number $l$ is demonstrated in Fig.\ref{width}(c). One can see that larger topological charge numbers destroy the spatial resolution, {\it i.e.}, the larger the OAM number $l$ is, the bigger the value of $a_r$ and the wider the localization widths are. This is understood by the fact that when the OAM number $l$ increases to larger numbers, the dark hollow center is increased in size, as indicated by Eq.(\ref{VO}). For example when $l=5$ and for $\kappa = 10$, $a_r$ becomes 1.39$\mu$m($2.89\lambda_c$), in contrast to the best case where we achieved only $a_r=4$nm(0.0083$\lambda_c$) for $l=1$. Therefore, in what follows we take $l=1$ in order to get the best results.

\subsection{Influence of the Rydberg shift $s$}
Because of the position-dependent nature of the control field, the Rydberg shift $s(r)$ is difficult to be completely compensated by the detuning $\Delta_c$, {\it i.e.} a perfect antiblockade condition by using an appropriate $\Delta_c$ that compensates $s(r)$ at every position is impossible.  
Once $\Delta_c-s(r)\neq 0$ and considering $I_c\ll I_p$ (around $r\approx0$), $\sigma_{rr}$ can be approximated as
\begin{equation}
 \sigma_{rr}(r) = \frac{1}{1 +[\Delta_c-s(r)]^2/w(r)^2}.
 \label{srr1}   
\end{equation}
The resolution $a_r$ then can be obtained when $|\Delta_c-s(r)|=w(r)$ corresponding to $ \sigma_{rr}=1/2$.
Luckily, as observed in Fig.\ref{Rbedge}b of appendix A, we see that $s(r_j)$ can be approximately a constant when the localized atom $j$ is placed within a short displacement $r_j$ to the beam core, {\it i.e.} if $r_j\ll R_b$. Hence adopting the way of so-called {\it partial antiblockade} in which we let the detuning $\Delta_c$ compensate the Rydberg interaction $s(r_j=0)$ exactly at the vortex core, is easy to implement practically. While if $r_j>0$, one obtains $\Delta_c-s(r_j)\neq0$ leading to a fast decrease of steady population.

\begin{figure}
\centering
\includegraphics[width=0.7\textwidth]{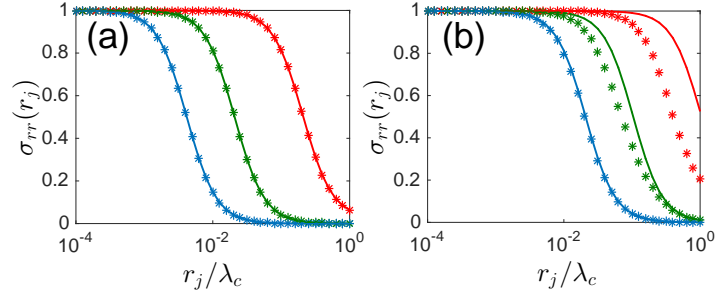} %
\caption{The influence of $s$ on the steady Rydberg population $\sigma_{rr}(r_j)$ under different beam waists (a) $W_0=1\mu$m and (b) $W_0=5\mu$m. Star points and solid curves are separately solved by considering a partial antiblockade $\Delta_c=s(r_j=0)$ and a perfect antiblockade $\Delta_c=s(r_j)$.  Plots corresponding to $\kappa=10,100,500$ are denoted by red, green and blue curves, respectively.}
\label{SrrW5}
\end{figure}

In Figure \ref{SrrW5}(a-b) we numerically compare the results for the cases of perfect antiblockade by $\Delta_c = s(r_j)$ (solid, the same as in Section 3.1) and the partial antiblockade (stars) featured by $\Delta_c = s(r_j=0)$. Comparing Figs. \ref{SrrW5}(a) and (b) shows that one can get a better localization when $W_0 = 1\mu$m. However, in this case, the partial antiblockade gives rise to the same results as the perfect antiblockade (see Fig.\ref{SrrW5}(a)), since the Rydberg interaction $s$ is approximately constant and can be compensated by a specific detuning $\Delta_c$ at every position. Therefore
the degree of localization can be well kept by an easier partial antiblockade condition, achieving the same FWHM e.g. $a_r \approx 4.0$nm for $\kappa=500$.

However, for a larger $W_0$ (e.g., $W_0=5\mu$m) it is insufficient to compensate $s(r_j)$ by a constant $\Delta_c$ that equals to $s(r_j=0)$, since the distortion and shrink of the blockade sphere caused by a wider beam waist $W_0$ can bring a significant variation to $s(r_j)$ making it spatial-dependent. Hence, we follow the partial antiblockade condition by letting $\Delta_c=s(r_j=0)$. In this way we find that once $r_j>0$, the steady Rydberg population $\sigma_{rr}(r_j)$ reveals a faster fall due to the imperfect compensation of $s$ beyond the localized point $r_j=0$. One finally obtains a narrower FWHM as compared to the case of perfect antiblockade, see Fig.\ref{SrrW5}(b).

Therefore, thanks to the hollow core laser beam we have realized an efficient scheme for ultraprecise 2D Rydberg localization. Remarkably, once the partial antiblockade condition $\Delta_c=s(r_j=0)$ is fulfilled, the Rydberg-Rydberg interaction counter-intuitively yields a better spatial resolution. This condition would be easier to be carried out in experiments.

\section{Longitudinal super localization}

\begin{figure}
\centering
\includegraphics[width=0.7\textwidth]{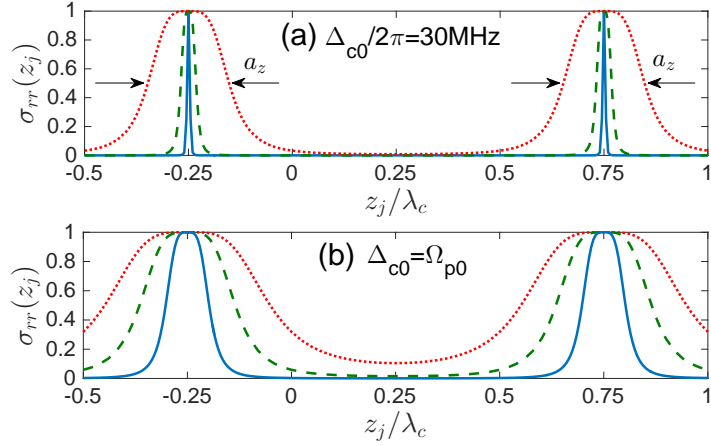} %
\caption{The periodic steady Rydberg distribution $\sigma_{rr}(z_j)$ along $z$ axis for $\kappa=10$(red-dotted), $100$(green-dashed), $500$ (blue-solid), under the peak-peak amplitudes (a) $\Delta_{c0}/2\pi=30$MHz (fixed) and (b) $\Delta_{c0}=\Omega_{p0}$ (tunable). Remember $\kappa = \Omega_{c0}/\Omega_{p0}$ and $\Omega_{c0}/2\pi=80$MHz. $a_z$ stands for the full-width at half maximal peak, {\it i.e.} $\sigma_{rr} \approx 0.5$ and $\lambda_c=480$nm is considered.}
\label{zwidth}
\end{figure}
 
Getting rid of using a SW optical Rabi frequency that may add to the complexity of our protocol, in the following we implement a spatial modulation to the detuning $\Delta_c(z)$ that is directly connected to $|r\rangle$. This modulation is enabled by the ac Stark effect from an external electric field to induce a periodic energy shift of $|r\rangle$ \cite{PRL083604}, taking the form of
\begin{equation}
    \Delta_c(z) = \Delta_{c0}\sin(\frac{2\pi}{\lambda_c} z)+\delta 
\label{Dz},
\end{equation}
with its peak-peak amplitude $\Delta_{c0}$ and an extra frequency shift $\delta$.
In such a case, the partial antiblockade condition changes to $\Delta_c = s(r_j=0,z_j = 3\lambda_c/4)$ at the localized point, giving
\begin{equation}
    \delta-\Delta_{c0}=s_0,
    \label{anti}
\end{equation}
 where $s_0= s(r_j = 0,z_j =3\lambda_c/4 )$. Once Eq.(\ref{anti}) is violated, {\it i.e.}, $\delta-\Delta_{c0}\neq s_0$, both the precision and spatial resolution will be reduced significantly, as numerically demonstrated in Fig.\ref{threeplot}. The
periodicity of the SW function $\Delta_c(z)$ allows atoms to be localized at $z_j = (3/4+n)\lambda_c$ [$n\in integers $]. Note that we have assumed $n=0$ in the simulations.

Based on the findings in Fig.\ref{Rbedge}(c) of Appendix A, it is safe to assume $s(r_j=0,z_j)$ to be a constant as $s(z_j)$ can be well kept with negligible oscillations. In this case, by following the equality of $\Delta_{c}(z)-s_0=w$, the FWHM of $\sigma_{rr}(z_j)$ can be solved analytically, leading to

\begin{equation}
a_z=[\frac{1}{2}-\arcsin(1-\frac{w}{\Delta_{c0}})/\pi]\lambda_c.
\label{az}    
\end{equation}

Figure \ref{zwidth} shows numerical simulations for the population distribution $\sigma_{rr}(z_j)$ against $z_j$. The numerical results for $a_z$ are in a good agreement with the analytical solutions given by Eq.(\ref{az}). Increasing $\kappa$ from 10 to 500 reduces $a_z$ significantly, yielding a tighter longitudinal confinement. Specifically, for a large control field detuning $\Delta_{c0}/2\pi=30$MHz and $\kappa=500$, the localization resolution can be enhanced, reaching $a_z=0.0046\lambda_c\approx2.2$nm (see Fig.\ref{zwidth}(a) (blue-solid)). 
However, if the peak-peak amplitude $\Delta_{c0}$ is set to be orders of magnitude smaller, {\it e.g.} $\Delta_{c0}=\Omega_{p0}$ as in Fig.\ref{zwidth}(b), it exhibits a dramatic broadening of $a_z$ due to $a_z\propto \Delta_{c0}^{-1}$, as featured by Eq.(\ref{az}).

\section{Experimental feasibility}

\subsection{Ultrahigh-precision 3D localization}
Our localization protocol benefits greatly not only from a hollow-core vortex beam which enables the confinement of atoms in a single site with a 100$\%$ detection probability, but also from the presence of Rydberg-Rydberg interaction which efficiently speeds up the fall of probability in space, finally improving the transverse localization resolution towards a subwavelength domain. In addition, applying a spatially-modulated two-photon detuning instead of an SW laser, enables us for precisely localizing atom along the longitudinal direction.

\begin{figure*}
\centering
\includegraphics[width=4.9in,height=2.6in]{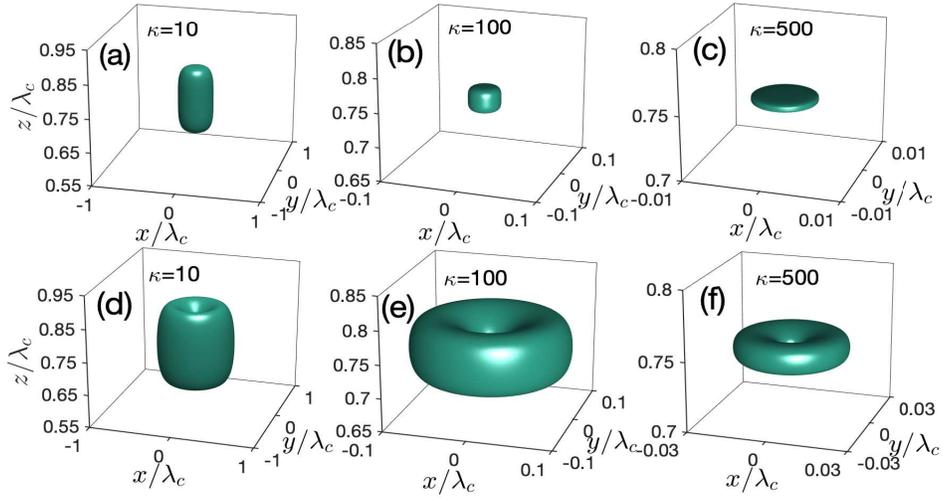}
\caption{(color online). Isosurface plots of Rydberg excitation probability at half maxima of $\sigma_{rr}$, versus $(x,y,z)$ for different $\kappa$ values under the case of (a-c) partial antiblockade condition: $\delta-\Delta_{c0}=s_0$; as well as in (d-f) that the partial antiblockade breaks by letting $\delta-\Delta_{c0}=2s_0$. Other relevant parameters are described detailedly in the text.}
\label{threeplot}
\end{figure*}

Before proceeding, we now numerically estimate the relevant experimental parameters for verifying the practical implementation of our protocol. 
In the calculations, we consider ground $^{87}\text{Rb}$ atoms with energy levels $(|g\rangle,|e\rangle,|r\rangle)=(|5s_{1/2}\rangle,|5p_{3/2}\rangle,|60s_{1/2}\rangle)$ excited by a two-photon process at the temperature of $20\mu$K \cite{193603}. The upper transition of $|e\rangle\to|r\rangle$ is played by a vortex LG beam with $(\Omega_{c0},W_0,\lambda_c)=(2\pi\times80\text{MHz},1\mu \text{m},480\text{nm})$ for the peak intensity, the beam waist and the wavelength, respectively. The lower transition from $|g\rangle$ to $|e\rangle$ is characterized by the continuous probe field with the Rabi frequency $\Omega_{p0}= \Omega_{c0}/\kappa$ and the wavelength 780nm. We introduced a tunable $\kappa$ which is manipulated by changing $\Omega_{p0}$, when $\Omega_{c0}/2\pi=80$MHz is fixed. In general, we take $\Omega_{p0}=2\pi\times(8.0,0.8,0.16)$MHz providing $\kappa=(10,100,500)$.
The average atom density is $\rho=6\times10^8$mm$^{-3}$, and the 
vdWs coefficient of state $|60s_{1/2}\rangle$ is $C_6/2\pi = 140$GHz$\mu \text{m}^6$\cite{C6}. The dissipation is dominated by a fast decay from the middle excited state $|e\rangle$, given by $\Gamma_e/2\pi = 6.05$MHz \cite{193603} and other spontaneous decays are $\Gamma_{r,g}=0$, leading to the dephasing rates $\gamma = \gamma_{er}=\gamma_{ge}=\Gamma_e/2$.


Besides, the two-photon detuning $\Delta_{c}(z)$ (note that $\Delta_p = 0$) is modulated as a sinusoidal function with its amplitude $\Delta_{c0}/2\pi= 30$MHz and periodicity $\lambda_c$. Importantly, the shifted energy $\delta$ should be decided accurately according to the partial antiblockade condition Eq.(\ref{anti}), leading to $\delta = \Delta_{c0}+s_0 = 2\pi\times(37.77,31.15,30.063)$MHz for $\kappa = (10,100,500)$, where $s_0$ is numerically estimated from solving an integration equation (\ref{s_0}) [see Appendix] at $r_j=0,z_j = 3\lambda_c/4$. If $\kappa\gg1$, $\delta$ is closing to the value of $\Delta_{c0}$ due to $s_0/\gamma\to 0 $. Therefore, in a real implementation in order to obtain an accurate $\delta$ for an optimal localization, one needs to scan the shifted frequency $\delta$ around $\Delta_{c0}$. A brief discussion for the  determination of the shifted frequency $\delta$ is presented in Appendix B.

Our final results for the 3D Rydberg atom localization are summarized in Fig.\ref{threeplot} with all practical parameters, where a global isosurface at half maxima of $\sigma_{rr}$ is represented. Panels (a-c) show the partial antiblockade case at the localized point: $\Delta_c = s(r_j=0,z_j = 3\lambda_c/4)$, leading to an accurate relation $\delta-\Delta_{c0}=s_0$. The plots demonstrate a visible localization image with the spatial resolution which can be optimized for a larger $\kappa$. However, once this accurate relation is violated {\it e.g.} $\delta-\Delta_{c0}=2s_0$ as indicated in (d-f), the FWHM of $\sigma_{rr}(\boldsymbol{r})$ experiences a dramatic broadening while decreasing its peak value. This confirms our theoretical predictions that the sensitive partial antiblockade condition at the localized point is very important for realizing a high-quality atom localization, especially in the presence of strong Rydberg-Rydberg interactions. A slight shift of the partial antiblockade condition would cause the breakdown of our localization scheme.

\subsection{Selection of the parameter $\kappa$}

\begin{figure}
\centering
\includegraphics[width=0.7\textwidth]{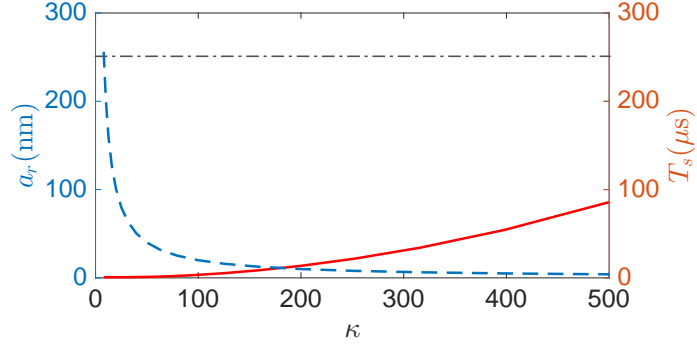}
\caption{The FWHM $a_r$(blue-dashed) and the steady time $T_s$(red-solid)  versus the change of the ratio $\kappa$. For $|r\rangle=|60s_{1/2}\rangle$ the Rydberg-state lifetime is 252$\mu$s as denoted by the dot-dashed line.}
\label{fig_kappa}
\end{figure}

As discussed in section 3, an ultra-precise localization relies on a sufficiently large $\kappa$[$\approx 500$].
However, the duration time $T_s$ for reaching a steady localization is typically inversely proportional to the absolute values of Rabi frequencies $\Omega_{c0}$ and $\Omega_{p0}$. As $\kappa=\Omega_{c0}/\Omega_{p0}$ ($\Omega_{c0}$ is fixed), $T_s$ is positively associated with $\kappa$. That means a large $\kappa$ leads to a sufficiently long time for the system to be stationary. 
In reality, $T_s$ is also limited by the Rydberg lifetime typically $T_s\ll1/\Gamma_r$ is required. For that purpose a proper $\kappa$ is mostly favored. {\it e.g.} by numerically solving Bloch Eqs.(\ref{Bloch_eqs}), we find when $\kappa=500$ the steady time $T_s$ reaches as high as 86$\mu$s which is comparable to the lifetime.

 In order to get an optimal $\kappa$ value, Fig.\ref{fig_kappa} shows the opposite dependence of the width $a_{r}$ and time $T_s$ on $\kappa$ that confirms our analytical predictions. In an experiment, the parameter $\kappa$ should be chosen properly by considering the precision of localization as well as the time duration to reach steady state at the same time. 
For example, if $\kappa=180$, it gives $T_s=11\mu$s, which is 23 times shorter than the lifetime of a Rydberg state. So this result can provide enough time for measuring the steady Rydberg probability due to its localization. Also, for $\kappa=180$ the spatial resolution is $a_r=11$nm ensuring a nanoscale-level localization precision. Therefore optimizing $\kappa$ is very important for the success of ultra-precise localization. In general replacing with a Rydberg state of higher principal quantum number can help to enhance $\kappa$, giving rise to a higher localization precision.

\section{Scheme stability}

\subsection{Laser intensity noise}

\begin{figure}
\centering
\includegraphics[width=0.7\textwidth]{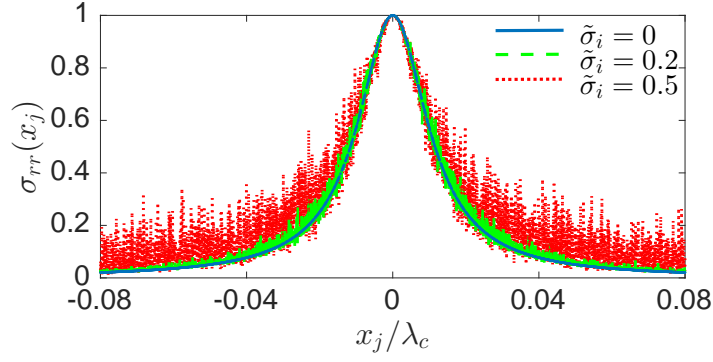}
\caption{The position-dependent steady Rydberg population $\sigma_{rr}(x)$ along $x$ axis under random laser intensity noise given by $\Omega_{c0}'(\mathbf{x})$.
The standard deviation of  $\Omega_{c0}'(\mathbf{x})$ is exemplified as  $\tilde{\sigma}_i=0.5\Omega_{c0}$(red-dotted), $\tilde{\sigma}_i=0.2\Omega_{c0}$(green-dashed) and $\tilde{\sigma}_i=0$(blue-solid). Each curve of $\tilde{\sigma}_i=(0.2,0.5)\Omega_{c0}$ is obtained by averaging over ten simulation trajectories and $\tilde{\sigma}_i=0$ is for the case with no noise. Here $\kappa=180$, $\Omega_{c0}/2\pi=80$MHz and the beam waist $W_0=1\mu$m.}
\label{noise}
\end{figure}

In order to explore the robustness of our scheme against external perturbations, we introduce now a position-dependent random intensity noise to the peak amplitude $\Omega_{c0}$ of the LG beam. In this case, the amplitude $\Omega_{c0}$ turns to be  position-dependent, defined by
\begin{equation}
\Omega_{c0}(\mathbf{r})=\Omega_{c0}^s+\Omega_{c0}'(\mathbf{r})
\end{equation}
with $\Omega_{c0}^s$ being the unperturbed laser amplitude. Here the random noise term $\Omega_{c0}'(\mathbf{r})$ is simulated as a normal distribution, which has a zero expectation value and a standard deviation $\tilde{\sigma_i}$. We only pay attention to the steady Rydberg distribution along $x$ axis due to the symmetry of system, in order to measure its variations under the effect of random fluctuations $\Omega_{c0}'(x)$. For reducing the uncertainty during single measurement, each plot is simulated by averaging over ten-times outputs.

Figure \ref{noise} demonstrates that suffering from the influence of the intensity noise, the steady Rydberg population $\sigma_{rr}(x_j)$ starts fluctuating and broadening. When the noise amplitude characterized by $\tilde{\sigma}_i$ is relatively small, {\it e.g.} $\tilde{\sigma}_i=0.2\Omega_{c0}$, the population distribution is quite steady closing to the case of $\tilde{\sigma}_i=0.0$ (no noise). Neverthless, if $\tilde{\sigma}_i$ is increased to be 0.5$\Omega_{c0}$, $\sigma_{rr}(x_j)$ reveals a strong fluctuation with its resolution (half-width at $\sigma_{rr}=0.5$) becoming worse. To our knowledge the realistic intensity noise can be suppressed to a very weak level under current experimental technique, which strongly supports the robustness of our localization protocol towards a new ultra-precision standard.

\subsection{Frequency noise}

\begin{figure}
\centering
\includegraphics[width=0.7\textwidth]{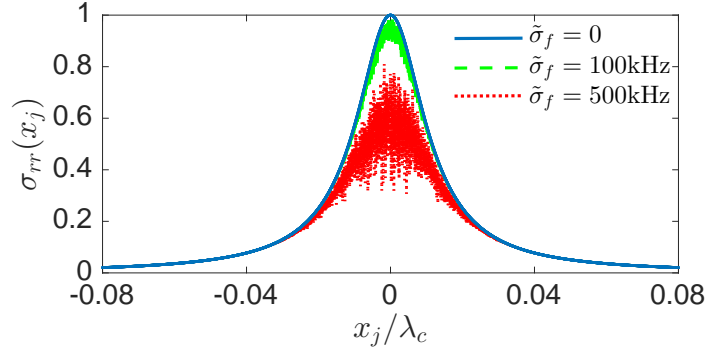}
\caption{The steady Rydberg population $\sigma_{rr}(x_j)$ {\it vs} the transverse direction $x_j$ suffering from a random noise of frequency shift $\delta_f$ in the periodically-modified detuning $\Delta_c(z)$. The standard deviation of the random noise is given by  $\tilde{\sigma}_f=500$kHz(red-dotted), $\tilde{\sigma}_f=100$kHz(green-dashed) and $\tilde{\sigma}_f=0$(blue-solid), where $\tilde{\sigma}_f=0$ means no frequency noise.
Here $s_0=\delta-\Delta_{c0}=2\pi\times 0.42$MHz stands for the Rydberg interaction at the localized points. Other parameters are the same with Fig.\ref{noise}.
}
\label{f_noise}
\end{figure}

Next, we explore the effect of frequency fluctuations to show the robust stability of our scheme.  A random frequency noise mainly contributed by the unavoidable laser-induced ac Stark effect \cite{PRAp014019,PRAp014031} is introduced to the frequency shift $\delta$ of the level detuning $\Delta_c(z)$, with a normal distribution and a small deviation $\tilde{\sigma}_f$. Applying additional compensation laser fields may reduce this effect \cite{Haffner:03}. Here the deviation $\tilde{\sigma}_f$ is modified to be with respect to $s_0$ since $s_0=\delta-\Delta_{c0}$ is small and very sensitive to $\delta$. As observed in Fig.\ref{f_noise}, it is clear that the steady Rydberg population $\sigma_{rr}(x_j)$ sensitively depends on the deviation strength because of the partial antiblockade condition $\delta-\Delta_{c0}=s_0$. If perturbed by a frequency noise $\tilde{\sigma}_f\neq0$ it leads to a significant influence on $\sigma_{rr}(x_j)$. Similar results are confirmed by Fig.\ref{threeplot}.

However, different from Figure \ref{noise}, the frequency noise would cause a dominant variation of the population in the vicinity of the beam core. As moving far away from the core, the effect of frequency noise becomes negligible. The reason can be understood by using Eq.\ref{sigmarr}, which can be re-expressed as
\begin{equation}
    \sigma_{rr}=\frac{1}{1+\eta+\frac{(\gamma^2+2I_p)}{(I_p+I_c)\Omega_p}(\Delta_c-s)^2}.
    \label{sigmarrn}
\end{equation}
Far from the beam core, the second term $\eta=I_c/I_p$ in the denominator is much larger than the third term $\frac{(\gamma^2+2I_p)}{(I_p+I_c)\Omega_p}(\Delta_c-s)^2$. In this case, the influence of the frequency noise with respect to $\sim (\Delta_c-s)^2$ becomes negligible. On the other hand, if the atom is placed in the vicinity of the beam core, $\eta=I_c/I_p$ goes to zero. The frequency noise then brings a remarkable fluctuation to $\sigma_{rr}$, and reduces the steady population probability.

\section{Concluding Remarks}
In conclusion, we have proposed a robust protocol for localizing the highly-excited Rydberg atoms. The periodicity of the SW field in earlier schemes was an obstacle for detecting atoms in single excitation regions. We have overcome this obstacle by applying an optical vortex, enabling a super transverse localization of Rydberg atoms solely in the vicinity of the vortex core and with a resolution down to the nanometer scale. The presence of the Rydberg-Rydberg interaction also yields a better localization when it is partially compensated by a suitable detuning. We have also demonstrated that a 3D localization is possible when applying simultaneously a vortex beam and an auxiliary SW modulation to the two-photon detuning. The SW modulation of the detuning provides a longitudinal confinement, while the vortex field localizes Rydberg atoms transversely. The vortex based approach has unique advantages that may be especially useful for Rydberg quantum computation in a nanometer-scale level, in which its robustness to the laser intensity noise revealed will offer special applications for the high-precision operation of a stable quantum logic gate.

\section*{Appendix A:Rydberg-Rydberg shifted energy $s(\boldsymbol{r})$}

\begin{figure*}
\centering
\includegraphics[width=0.7\textwidth]{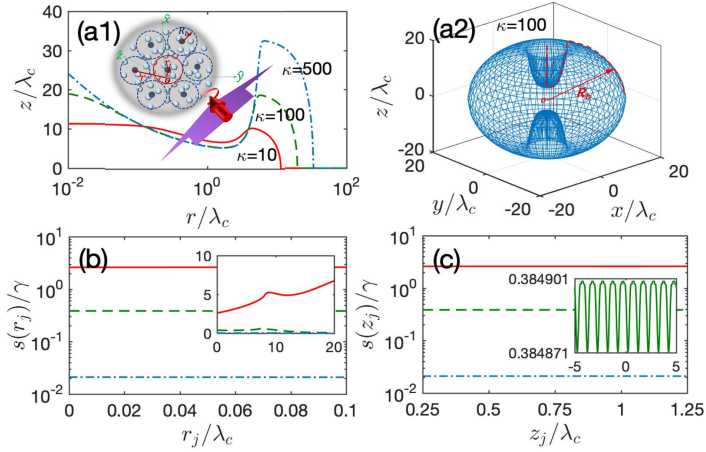} %
\caption{(a1) The position-dependent blockade boundary for atom $j$ at $r_j= 0$, $z_j=3\lambda_c/4$, characterized by an anisotropic blockade radius $R_b$ in the $(r,z)$ space for $\kappa$=10(red solid), 100(green-dashed), 500(blue dash-dotted) respectively. Inset: schematic diagram of superatom ensembles illuminated by a localized LG field $\Omega_c(r,\phi)$, as well as a non-localized TW field $\Omega_p$. (a2) A 3D visible plot of the anisotropic blockade sphere with blockade radius $R_b$ and $\kappa =100$. Note that (a1) is a cross-section of the blockade sphere, as denoted by the red-dashed curve. (b) The Rydberg shifted energy $s(r_j)$ at $z_j =3\lambda_c/4 $ versus $r_j\in[0,0.1]\lambda_c$. A wider range of $r_j\in[0,20]\lambda_c$ is given in the inset, where $s(r_j)$ shows a slight change with $r_j$. Similarly to (b), (c) represents the relationship between $s(z_j)$ and $z_j\in[0.25,1.25]\lambda_c$ at $r_j=0$. Inset denotes a detailed image for $s(z_j)$, where tiny oscillations are preserved, stemming from the SW periodicity of $\Delta_c(z)$.}
\label{Rbedge}
\end{figure*}

The position-dependent shifted energy $s(\boldsymbol{r})$ can be solved for numerically. In the frame of cylindrical coordinates, from Eq. (8) $s(r_j,z_j)$ of the $j$th atom contributed by other surrounding $m$ atoms has a reduced dual-integration form, described by
\begin{equation}
 s(r_j,z_j)=2\pi C_{6}\int_0^{\infty}\int_{{-\infty}}^{\infty}\frac{\bar{\sigma}_{rr}\rho \chi}{[(r-r_j)^2+(z-z_j)^2]^{3}} rdzdr,
 \label{s_chi}
\end{equation}
where the azimuthal angle $\phi$ has been dropped out due to the symmetry. An adjustable coefficient $\chi$ 
\begin{equation}
\chi=\begin{cases}
0 & r^{2}+z^{2}<R_{b}^{2}\\
1 & r^{2}+z^{2}\geq R_{b}^{2}
\end{cases}
\end{equation}
is introduced to control the interaction strength of adjacent atoms $m$.
Hence, if the adjacent atom $m$ is placed inside the blockade sphere preventing all Rydberg excitations, then $\chi=0$ and $s=0$; otherwise $\chi=1$, leading to $s\neq0$. The excited atom $m$ will induce a finite Rydberg shift to the atom $j$ at $(r_j,z_j)$.
In the calculations, the entire integration regime contains a computational lattice with  $10^4\times10^4$ points in $(r,z)$-directions, where the computational lengths and the lattice spacing along each dimension are  $L_{r}=L_{z}=100\lambda_c\sim48\mu$m and $\delta r=\delta z=0.01\lambda_c\sim4.8$nm.

In Fig. \ref{Rbedge}(a1) we study the position-dependent blockade radius $R_b(r,z)$ for $\kappa = 10,100,500$. Explicitly, when the localized atom $j$ is placed at the core of the LG field, $R_b$ is essentially anisotropic and increases with $\kappa$. The reason is, that near the beam core, where $I_c\approx0$, $R_b$ is inversely proportional to $I_p$. However, a common dip occurs at the beam waist around $r\approx1.5\lambda_c$, arising from the fact that at this point the intensity $I_c$ arrives at a same maximal value $I_{c0}$. Due to the dominant role played by the strong intensity $I_{c0}$, the blockade radius tends to be the same around $r\approx1.5\lambda_c$ no matter what $\kappa$ is. The inset shows a visual picture of a collection of superatom ensembles illuminated by the LG (red) and TW (amaranth) fields. A 3D plot of the anisotropic blockade sphere is given in Fig. \ref{Rbedge}(a2) to visualize the imaging, in which the blockade radius $R_b$ in three-dimension is stressed. Note that (a1) is only a part of cross-section boundaries as denoted by the red-dashed curve in the 3D blockade sphere.

Guided by the anisotropy of blockade radius, we further exploit the accumulated Rydberg shift $s(r_j,z_j)$ via the variation of positions, by considering $z_j=3\lambda_c/4$ (Fig. \ref{Rbedge}(b)) and $r_j=0$ (Fig. \ref{Rbedge}(c)), separately. Note that the atom $j$ is localized at $(r_j,z_j)=(0,(3/4\pm n)\lambda_c)$ with $n=0,1,2...$ denoting the periodic number coming from the SW modulation $\Delta_c(z)$. Here we choose $n=0$. Based on Fig. \ref{Rbedge}(b-c), generally speaking the value of $s(r_j,z_j=3\lambda_c/4)$ or $s(r_j=0,z_j)$ can be robustly preserved no matter how $\kappa$ is tuned, benefiting from the tiny localization regime around $r_j=0$ and $z_j=3\lambda_c/4$. For example Fig. \ref{Rbedge}(b) shows the little variation of the interaction $s$ within $r_j\leq48$nm.
That preservation property gives rise to a partial antiblockade relation by $\delta-\Delta_{c0}=s_0$ where $s_0$ means the shifted energy at the localized point, and $\delta$, $\Delta_{c0}$ are related to the modulation function $\Delta_c(z)$. Beyond the localization regime an insufficient preservation due to partial antiblockade effect can counter-intuitively speed up the fall of the excited-state probability, making the atom position confined within a narrower area, as indicated in SubSec. \rm{III}.B.

On the other hand, it is also confirmed that $s(r_j,z_j)$ significantly decreases for a larger $\kappa$. The reason is that, if $\kappa$ is large, the localized atoms placed around the core would suffer from a weaker interaction from other atoms due to the sufficient size of the blockade radius. An extensive plot in (b) supplementarily shows that the shifted energy $s(r_j)$ indeed changes slightly if the atom is placed far from the beam core, where $r_j$ is a few micrometers as comparable as $R_b$, especially for a small $\kappa$. At the same time, the shifted energy $s(z_j)$ keeps a long-range and stable preservation along the $z$ axis in every period [see inset of (c)].

\section*{Appendix B:The shifted frequency $\delta$}

\begin{figure}
\centering
\includegraphics[width=0.7\textwidth]{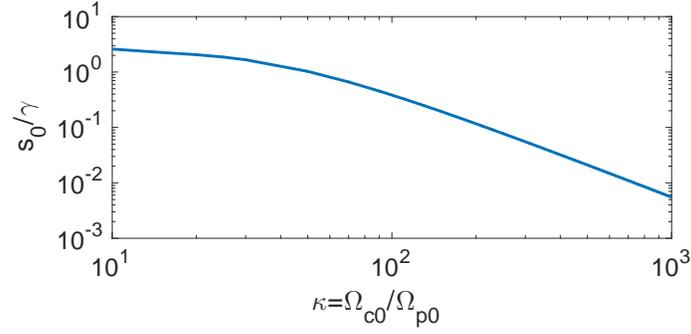} %
\caption{The shifted energy $s_0$ at the localized point $(r_j,z_j)=(0,3\lambda_c/4)$ versus the change of the ratio $\kappa$.}
\label{s0kappa}
\end{figure}

To realize a robust 3D localization, we reveal the importance of the partial blockade relation $\delta-\Delta_{c0}=s_0$ with $\Delta_{c0}$ being the peak-peak modulation amplitude. Here $\Delta_{c0}$ is arbitrarily chosen to be $\Delta_{c0}/2\pi=30$MHz. The shifted frequency $\delta$ should be determined by $\delta=s_0+\Delta_{c0}$. Here $s_0$ stands for the average Rydberg shift at the localized point $(r_j,z_j)=(0,3\lambda_c/4)$, which can be solved by rewriting Eq.(\ref{s_chi}) into

\begin{equation}
s_0=2\pi C_{6}\rho I_p\int_0^{\infty} r \int_{-\infty}^{\infty}\frac{\chi}{[(r-r_j)^2+(z-z_j)^2]^{3}B} dzdr,
\label{s_0}
\end{equation}
where $B=I_c(r)+4\pi R_b^3 \rho I_{p0}/3+\frac{(\gamma^2+2I_{p0})\Delta_c(z)^2}{I_{p0}+I_c(r)}$. If $s_0$ is numerically solved it is possible to apply the value $\delta=s_0+\Delta_{c0}$ for structuring the sinusoidal modulation $\Delta_c(z)$. Fig.\ref{s0kappa} plots the behavior of $s_0$ with respect to $\kappa$. It is clearly shown that $s_0$ decreases significantly as $\kappa$ grows, and $s_0$ also preserves a tiny value if $\kappa$ is very large. For example as $\kappa>100$, $s_0$ has entered the regime below $\sim0.1\gamma$ or smaller. Owing to the use of a weak probe field $\Omega_{p0}$, the poor Rydberg excitation probability can lead to a smaller Rydberg shift $s_0$.

However, although $s_0$ is very small, the chosen  $\delta$ should be sensitive to it. Once the relation of $\delta = s_0+\Delta_{c0}$ breaks, a dramatic broadening of FWHM occurs, that can greatly reduce the spatial resolution of the localization [see Fig.5(d-f)]. In experiment one needs to scan the frequency $\delta$ very precisely around $\Delta_{c0}$ to improve the localization quality. For example, we numerically obtain the values of $\delta/2\pi =(37.77,31.15,30.063)$MHz for $\kappa = (10,100,500)$, confirming that $\delta$ becomes very close to $\Delta_{c0}=30$MHz when $\kappa$ is sufficiently large.

\section*{Acknowledgments}
J.Q. acknowledges the useful discussions from Bing Chen. This work was supported by the National Natural Science Foundation of China under Grants No. 11474094 and No. 11104076 and the Science and Technology Commission of Shanghai Municipality under Grant No. 18ZR1412800 for J.Q., by the STSM Grants for T.K. from COST Action CA16221, and has received funding from European Social Fund (Project No. 09.3.3-LMT-K-712-19-0031) under grant agreement with the Research Council of Lithuania (LMTLT) for H.R.H.

\section*{Disclosures}
The authors declare that there are no conflicts of interest related to this article.


\bibliography{localization}

\end{document}